\documentclass[aps,prd,reprint,superscriptaddress,nofootinbib]{revtex4-2}

\usepackage[T1]{fontenc}
\usepackage{amsmath,amssymb,amsfonts,mathtools}
\usepackage{bm}
\usepackage{graphicx}
\graphicspath{{./}}
\usepackage{siunitx}
\usepackage{microtype}
\usepackage[colorlinks=true,linkcolor=blue,citecolor=blue,urlcolor=blue]{hyperref}

\allowdisplaybreaks
\DeclareSIUnit\dalton{Da}
\DeclareSIUnit\parsec{pc}

\newcommand{\Tr}{\operatorname{Tr}}
\newcommand{\dd}{\mathrm{d}}
\newcommand{\ii}{\mathrm{i}}
\newcommand{\ee}{\mathrm{e}}
\newcommand{\kk}{\bm{k}}
\newcommand{\xx}{\bm{x}}

\newcommand{\nn}{\hat{\bm{k}}}
\newcommand{\Mpl}{m_{\mathrm P}}
\newcommand{\TT}{\mathrm{TT}}
\newcommand{\GH}{G^{\mathrm H}}
\newcommand{\order}{\mathcal{O}}
\newcommand{\Dcoh}{\mathcal{D}}

\begin{document}

\title{Graviton-induced which-path decoherence in matter-wave interferometry}
\author{Hiroki Matsui}
\affiliation{Osaka Central Advanced Mathematical Institute (OCAMI), Osaka Metropolitan University, Osaka 558-8585, Japan}
\date{\today}

\begin{abstract}
We derive which-path decoherence for matter-wave interferometry that arises from tracing out the gravitons of linearized quantum gravity. Because the gravitons are driven linearly by the matter source, the branch-dependent evolution can be solved exactly, and the reduced coherence is given by the characteristic function of the initial graviton state, evaluated at the difference of the branch-induced field displacements. In vacuum, the decoherence exponent equals half the mean number of gravitons radiated by the difference source. For a smooth Gaussian trajectory, it reduces to $\Gamma_{\rm vac}=(8/15)\,Gm^2d^4/(\hbar c^5\tau^4)$, ranging from $10^{-89}$ to $10^{-61}$ across representative matter-wave platforms. For a general squeezed graviton vacuum, we obtain an exact expression in which squeezing either suppresses or enhances the vacuum response, depending on the squeezing phase. For the strongly squeezed state produced by inflation, this expression reduces to $\Gamma_{\rm inf}=(\pi/20)\,\Omega_{\rm inf}(md^2H_0/\hbar)^2$ and reaches at most $\sim10^{-27}$ for optimistic parameters. Hence, the which-path decoherence induced by radiative gravitons is therefore negligible in current matter-wave interferometers.
\end{abstract}

\maketitle

\section{Introduction}
\label{sec:intro}

Whether the gravitational field must be quantized and whether its quantum nature leaves any observable imprint on laboratory systems remain central open questions in fundamental physics. A minimal and theoretically clean setting in which a quantized field produces a distinctive effect is \emph{gravitational decoherence}, an instance of environment-induced decoherence~\cite{Joos:1984uk,Breuer:2002pc,Zurek:2003zz}: a massive body prepared in a superposition of two semiclassical histories sources branch-dependent gravitational fields, and the associated radiative graviton states can carry which-path information, thereby reducing the interference visibility after recombination. Within the low-energy effective field theory of gravitons coupled to matter, this mechanism has been developed into a quantitative framework for describing the gravitons associated with superposed sources in flat spacetime~\cite{Riedel:2013yca} and at Killing horizons~\cite{Danielson:2022tdw,Danielson:2022sga,Danielson:2024yru}. Photon-bremsstrahlung decoherence is the electromagnetic analogue~\cite{Baym:2009zu}; Ref.~\cite{Carney:2018ofe} provides a broader review. The Feynman--Vernon formalism~\cite{Feynman:1963fq} underpins influence-functional and master-equation studies of graviton-induced decoherence and state-dependent graviton noise in detectors~\cite{Blencowe:2012mp,Anastopoulos:2013zya,Oniga:2015lro,Parikh:2020nrd,Kanno:2020usf,Toros:2020krn,Parikh:2020kfh,Parikh:2020fhy,Kanno:2021gpt}. Together with analyses of gravitational radiation from quantum systems~\cite{Ford:1982wu} and nonclassical primordial gravitational waves~\cite{Grishchuk:1990bj,Kanno:2018cuk,Kanno:2024gjt}, these studies complement this picture. Astrophysical studies likewise consider coherent and squeezed graviton states and propose graviton statistics as probes of primordial nonclassicality~\cite{Kanno:2025how,Dorlis:2025zzz,Dorlis:2025amf,Kanno:2025fpz}

Matter-wave interferometry thus involves two distinct gravitational effects. The first is a branch-dependent phase underlying proposals to entangle mesoscopic masses via gravity~\cite{Bose:2017nin,Marletto:2017kzi}, including optomechanical schemes~\cite{Matsumura:2020law,Miki:2024qcz,Fujita:2023pia}; tracing out environmental matter in Newtonian models can also cause decoherence~\cite{Miki:2020hvg,Takeda:2026ujz}. The second is path--field entanglement which suppresses visibility when gravitons are traced out, while influence-functional analyses characterize graviton noise in vacuum and squeezed states~\cite{Riedel:2013yca,Parikh:2020nrd,Kanno:2020usf,Parikh:2020kfh,Parikh:2020fhy,Kanno:2021gpt}. Here we treat this radiative effect exactly for a closed, conserved source using displacement operators, without $S$-matrix or Markovian approximations, and restrict attention to propagating gravitons in Minkowski spacetime.

In this work, we compute the which-path decoherence that results from tracing
out propagating gravitons in
linearized quantum gravity. The key simplification is that the gravitons are driven linearly by the matter stress tensor. Each branch of
the superposition therefore induces a coherent displacement of every
field mode, and the branch evolution can be solved exactly: the
reduced coherence equals the characteristic function of the initial
graviton state, evaluated at the difference of the branch-induced field
displacements. This formulation is exact for an arbitrary initial graviton state and cleanly separates a relative phase from a
loss of visibility. We evaluate the resulting decoherence for physically motivated initial states of the graviton field, namely the
Minkowski vacuum and the squeezed state produced by inflation.

For the vacuum, the decoherence exponent equals half the mean number of
gravitons radiated by the difference source. Reducing this to the nonrelativistic
quadrupole limit and adopting a smooth Gaussian trajectory model yields
the compact estimate $\Gamma_{\rm vac}=(8/15)\,Gm^2d^4/(\hbar
c^5\tau^4)$, which ranges from approximately $10^{-89}$ to $10^{-61}$ across
representative matter-wave platforms. For a general squeezed vacuum we
obtain an exact expression in which squeezing suppresses or enhances the
vacuum response. Specializing to
the primordial graviton background generated by
inflation, we find that the phase-sensitive contribution
averages out over the experimental frequency band, leaving the enhancement $\Gamma_{\rm inf}=(\pi/20)\,\Omega_{\rm
inf}(md^2H_0/\hbar)^2$, which reaches at most $\sim10^{-27}$ for
optimistic parameters. Although inflationary squeezing enhances the
decoherence exponent by approximately 40 orders of magnitude relative to the vacuum, the
decoherence remains far below any plausible detection threshold.

The remainder of this paper is organized as follows.
Section~\ref{sec:conventions} fixes our conventions for linearized
gravity and graviton quantization. Section~\ref{sec:evolution} solves the
branch-dependent evolution in closed form and derives the exact reduced
coherence. Section~\ref{sec:vacuum} evaluates the vacuum decoherence and
relates it to the radiated graviton number;
Sec.~\ref{sec:quadrupole} then reduces the master formula to the nonrelativistic quadrupole
limit. Section~\ref{sec:gaussian} applies this to the Gaussian trajectory
model, and Sec.~\ref{sec:squeezed} extends the analysis to squeezed and
inflationary graviton states. Section~\ref{sec:experiment} presents
numerical estimates for representative experiments, and
Sec.~\ref{sec:discussion} concludes.

\paragraph*{Conventions.}
We use the mostly-plus metric $\eta_{\mu\nu}=\mathrm{diag}(-1,+1,+1,+1)$. Greek indices denote spacetime components, and Latin indices denote spatial components. We set $\hbar=c=1$ through Sec.~\ref{sec:vacuum} and restore SI units beginning in Sec.~\ref{sec:quadrupole}. Our Fourier convention is
\begin{equation}
 \widetilde T^{\mu\nu}(\omega,\kk)
 =\int\dd^4x\,
 \ee^{\ii\omega t-\ii\kk\cdot\xx}\,T^{\mu\nu}(t,\xx),
 \label{eq:FTdef}
\end{equation}
and $\widetilde f(\omega)=\int\dd t\,\ee^{\ii\omega t}f(t)$ for time-dependent functions.

\section{Linearized gravity}
\label{sec:conventions}

This section summarizes standard results in linearized gravity to fix our conventions and normalizations. Readers familiar with graviton quantization may proceed directly to Sec.~\ref{sec:evolution}. The Einstein--Hilbert action and the metric expansion are
\begin{align}
 S_{\rm EH}&=\frac{2}{\kappa^2}\int\dd^4x\,\sqrt{-g}\,R,
 \label{eq:SEH}\\
 g_{\mu\nu}&=\eta_{\mu\nu}+\kappa\,h_{\mu\nu},
 \quad
 \kappa\equiv\sqrt{32\pi G},
 \label{eq:metricexp}
\end{align}
so that $h_{\mu\nu}$ has mass dimension one and a canonically normalized kinetic term.
Up to a total derivative, the quadratic action is the Fierz--Pauli form
\begin{align}
 S_{\rm EH}^{(2)}&=\int\dd^4x\,\Big[
 -\tfrac{1}{2}(\partial_\lambda h_{\mu\nu})(\partial^\lambda h^{\mu\nu})
 +(\partial_\mu h^{\mu\nu})(\partial^\lambda h_{\lambda\nu})
 \nonumber\\
 &\hspace{1.4cm}
 -(\partial_\mu h^{\mu\nu})\partial_\nu h
 +\tfrac{1}{2}(\partial_\lambda h)(\partial^\lambda h)\Big],
 \label{eq:FierzPauli}
\end{align}
where $h=\eta^{\mu\nu}h_{\mu\nu}$. This action is invariant to linear order under $h_{\mu\nu}\to h_{\mu\nu}+\partial_\mu\xi_\nu+\partial_\nu\xi_\mu$. The matter stress tensor is defined by variation with respect to the metric, so the linear coupling is
\begin{equation}
 S_{\rm int}=\frac{\kappa}{2}\int\dd^4x\,
 h_{\mu\nu}T^{\mu\nu}.
 \label{eq:Sint}
\end{equation}
For a conserved stress tensor, $\partial_\mu T^{\mu\nu}=0$, the coupling to physical on-shell graviton modes is gauge invariant. Restricting the gravitational field to its freely propagating helicity-two sector, we choose TT gauge and write the interaction-picture Hamiltonian as
\begin{equation}
 H_{\rm int}(t)
 =-\frac{\kappa}{2}\int\dd^3x\,
 \hat h^{\TT}_{ij}(t,\xx)\,T^{ij}(t,\xx).
 \label{eq:Hint}
\end{equation}

The graviton field is expanded as
\begin{align}
 \hat h^{\TT}_{ij}(x)
 &=\sum_{s=+,\times}\int\!\frac{\dd^3k}{(2\pi)^3}
 \frac{1}{\sqrt{2\omega_k}}
 \Big[
 \epsilon^s_{ij}(\nn)\,\hat a_s(\kk)\,
 \ee^{\ii(\kk\cdot\xx-\omega_k t)}
 \nonumber\\[-2pt]
 &\hspace{2.0cm}
 +\epsilon^{s*}_{ij}(\nn)\,\hat a_s^\dagger(\kk)\,
 \ee^{-\ii(\kk\cdot\xx-\omega_k t)}
 \Big],
 \label{eq:modeexp}
\end{align}
where $\omega_k=|\kk|$ and
\begin{equation}
 [\hat a_s(\kk),\hat a_{s'}^\dagger(\kk')]
 =(2\pi)^3\,\delta_{ss'}\,\delta^{(3)}(\kk-\kk').
 \label{eq:commutator}
\end{equation}
The polarization tensors satisfy
\begin{equation}
 \epsilon^s_{ii}=0,
 \quad
 \hat k_i\epsilon^s_{ij}=0,
 \quad
 \epsilon^s_{ij}\epsilon^{s'*}_{ij}=\delta_{ss'},
 \label{eq:polnorm}
\end{equation}
with completeness relation
\begin{equation}
 \sum_s\epsilon^{s}_{ij}(\nn)\,\epsilon^{s*}_{kl}(\nn)
 =\Lambda_{ij,kl}(\nn),
 \label{eq:polsum}
\end{equation}
where the TT projector is
\begin{align}
 \Lambda_{ij,kl}
 &=\tfrac{1}{2}\left(P_{ik}P_{jl}+P_{il}P_{jk}-P_{ij}P_{kl}\right),
 \label{eq:TTprojector}\\
 P_{ij}&=\delta_{ij}-\hat k_i\hat k_j.
 \label{eq:transverseprojector}
\end{align}
Our polarization tensors have unit norm, $\epsilon^s_{ij}\epsilon^{s*}_{ij}=1$, whereas some gravitational-wave references use tensors with norm $2$, $\epsilon^s_{ij}\epsilon^{s*}_{ij}=2$. All coefficients below follow the convention of Eq.~\eqref{eq:polnorm}. We also adopt, without loss of generality, the convention $\epsilon^s_{ij}(-\nn)=\epsilon^{s*}_{ij}(\nn)$, which is convenient when pairing opposite momenta in Sec.~\ref{sec:squeezed}.


\section{Evolution of matter and gravitons}
\label{sec:evolution}

\subsection{Physical setup}
\label{sec:setup}

Let $|1\rangle$ and $|2\rangle$ denote two narrow, approximately orthogonal semiclassical matter branches. At the initial time, we assume a product state of the branch degree of freedom and the radiative graviton modes,
\begin{equation}
\rho_{\rm in}
=\sum_{a,b=1}^2\rho^{\rm m}_{ab}\,|a\rangle\langle b|\otimes\rho_{\rm g},
\qquad
\sum_a\rho^{\rm m}_{aa}=1.
\label{eq:rhoin}
\end{equation}
An equal coherent superposition has $\rho^{\rm m}_{ab}=1/2$. We take $\rho_{\rm g}=|0\rangle\langle0|$ through Sec.~\ref{sec:gaussian} and $\rho_{\rm g}=|{\rm sq}\rangle\langle{\rm sq}|$ in Sec.~\ref{sec:squeezed}. The latter is a pure two-mode squeezed vacuum; Sec.~\ref{sec:inflation} specializes this state to the Bunch--Davies vacuum expressed in a late-time basis. The factorization in Eq.~\eqref{eq:rhoin} refers to the freely propagating TT gravitons.

Each branch $b=1,2$ is represented by a classical \emph{total} stress tensor $T_b^{\mu\nu}(x)$, and we define
\begin{equation}
\Delta T^{\mu\nu}\equiv T_1^{\mu\nu}-T_2^{\mu\nu}.
\label{eq:DeltaT}
\end{equation}
For the closed source model used below,
\begin{equation}
\partial_\mu T_b^{\mu\nu}=0,
\quad
\partial_\mu\Delta T^{\mu\nu}=0.
\label{eq:conservation}
\end{equation}
For a closed interferometric history, the branch difference is localized in time: the branches coincide before splitting and after recombination. Components of the apparatus that are common to the two branches cancel from $\Delta T^{\mu\nu}$.

\subsection{Closed-form evolution operator}

Because $H_{\rm int}(t)$ is linear in the creation and annihilation operators with $c$-number source coefficients, $[H_{\rm int}(t),H_{\rm int}(t')]$ is a $c$-number; all nested commutators therefore vanish, and the Magnus expansion terminates at second order \cite{Magnus:1954zz,Blanes:2008xlr}. The exact branch evolution is
\begin{align}
\hat U_b&=\ee^{\ii\Phi_b}\,\hat D[\alpha_b],
\nonumber\\
\hat D[\alpha_b]&\equiv
\exp\!\Big[\sum_s\!\int\!\frac{\dd^3k}{(2\pi)^3}
\big(\alpha_{b,s}\hat a_s^\dagger-\alpha_{b,s}^{*}\hat a_s\big)\Big].
\label{eq:Ub}
\end{align}
Equation~\eqref{eq:Ub} shows that a classical source coherently displaces every field oscillator, just as a classical current produces coherent radiation in electromagnetism. When the initial field is in the vacuum, each branch therefore produces a coherent state~\cite{Kanno:2025how}. All which-path information carried by the field is contained in the displacement amplitudes $\alpha_{b,s}(\kk)$, given by
\begin{equation}
\alpha_{b,s}(\kk)
=\frac{\ii\kappa}{2\sqrt{2\omega_k}}\,
\epsilon^{s*}_{ij}(\nn)\,
\widetilde T_b^{ij}(\omega_k,\kk).
\label{eq:alpha}
\end{equation}
Here $\hat U_b$ is the interaction-picture evolution operator for the complete protocol, evaluated at a readout time after the two branches have recombined. Accordingly, Eq.~\eqref{eq:alpha} contains the Fourier transform of the source over its entire history.
The second Magnus term gives the real retarded phase
\begin{align}
\Phi_b
&=\frac{\kappa^2}{8}\int\dd^4x\dd^4x'\;
T_b^{ij}(x)\,G^{R}_{ij,kl}(x-x')\,T_b^{kl}(x'),
\label{eq:Phib}
\end{align}
where $G^{R}_{ij,kl}(x-x')\equiv \ii\theta(t-t')[\hat h^{\TT}_{ij}(x),\hat h^{\TT}_{kl}(x')]$. Local renormalization terms implicit in Eq.~\eqref{eq:Phib} are assumed to be absorbed into the matter parameters; only finite branch-dependent phase differences are retained. This radiative phase is distinct from the static Newtonian phase generated by the constraint sector.

\subsection{Tracing out gravitons}

The evolution operator of the full system is
\begin{equation}
 \hat U=\sum_{b=1}^2|b\rangle\langle b|\otimes
 \ee^{\ii\Phi_b}\hat D[\alpha_b].
 \label{eq:controlledU}
\end{equation}
Let $\hat X[\alpha]$ denote the exponent in Eq.~\eqref{eq:Ub}; the commutator of two such exponents is the $c$-number
\begin{equation}
 [\hat X[\alpha],\hat X[\beta]]
 =\langle\beta,\alpha\rangle-\langle\alpha,\beta\rangle,
 \label{eq:Xcomm}
\end{equation}
with the mode-space inner product
\begin{equation}
 \langle\alpha,\beta\rangle
 \equiv\sum_s\int\frac{\dd^3k}{(2\pi)^3}\,
 \alpha_s^*(\kk)\,\beta_s(\kk).
 \label{eq:innerproduct}
\end{equation}
The Baker--Campbell--Hausdorff formula gives
\begin{align}
\begin{split}
 \hat D^\dagger[\alpha_2]\hat D[\alpha_1]
 &=\ee^{\ii\,\mathrm{Im}\langle\alpha_2,\alpha_1\rangle}\,
 \hat D[\Delta\alpha],\\
 \Delta\alpha &\equiv\alpha_1-\alpha_2.
 \label{eq:Dcomposition}
\end{split}
\end{align}
After the gravitons are traced out, the off-diagonal element of the reduced matter density matrix is multiplied by the coherence factor
\begin{align}
 \Dcoh \equiv \ee^{\ii(\Phi_1-\Phi_2)}\,
 \Tr\!\left[\rho_{\rm g}\,
 \hat D^{\dagger}[\alpha_2]\,\hat D[\alpha_1]\right].
 \label{eq:Dprecomp}
\end{align}
Using Eq.~\eqref{eq:Dcomposition}, we obtain
\begin{equation}
\Dcoh =\ee^{\,\ii\Delta\Phi_{\rm tot}}
\Tr\!\left[\rho_{\rm g}\hat D[\Delta\alpha]\right],
\label{eq:exactD}
\end{equation}
where $\Delta\Phi_{\rm tot}\equiv\Phi_1-\Phi_2+\mathrm{Im}\langle\alpha_2,\alpha_1\rangle$.
Equation~\eqref{eq:exactD} separates the effect of the graviton field into a relative phase $\Delta\Phi_{\rm tot}$ and a visibility factor $|\Tr\!\left[\rho_{\rm g}\hat D[\Delta\alpha]\right]|\le 1$. The corresponding decoherence exponent is
\begin{equation}
\Gamma\equiv-\ln|\Dcoh|
=-\ln\Big|\Tr\!\left[\rho_{\rm g}\hat D[\Delta\alpha]\right]\Big|.
 \label{eq:Gammadef}
\end{equation}
Thus, the surviving interference is governed by the overlap of the field states left behind by the two branches. When the two field states coincide, the interference fringes remain intact ($\Gamma=0$); when they are completely distinguishable, the fringes disappear ($\Gamma\to\infty$).

\section{Vacuum graviton decoherence}
\label{sec:vacuum}

For the graviton vacuum,
\begin{equation}
 \langle0|\hat D[\beta]|0\rangle
 =\exp\!\left[-\frac{1}{2}
 \sum_s\int\frac{\dd^3k}{(2\pi)^3}|\beta_s(\kk)|^2\right].
 \label{eq:vacchi}
\end{equation}
Equations~\eqref{eq:exactD} and \eqref{eq:alpha} then give
\begin{align}
 \Gamma_{\rm vac}
 &=\frac{1}{2}\sum_s\int\frac{\dd^3k}{(2\pi)^3}
 |\Delta\alpha_s(\kk)|^2
 \nonumber\\
 &=\frac{\kappa^2}{16}\sum_s
 \int\frac{\dd^3k}{(2\pi)^3\omega_k}
 \left|\epsilon^{s*}_{ij}(\nn)
 \Delta\widetilde T^{ij}(\omega_k,\kk)\right|^2.
 \label{eq:Gammavac0}
\end{align}
Substituting the polarization sum~\eqref{eq:polsum} and the relation $\kappa^2/16=2\pi G$, we obtain
\begin{equation}
 \Gamma_{\rm vac}
 =2\pi G\int\!\frac{\dd^3k}{(2\pi)^3\omega_k}\;
 \Lambda_{ij,kl}(\nn)\,
 \Delta\widetilde T^{ij}\,\Delta\widetilde T^{kl*}.
 \label{eq:masterTT}
\end{equation}
The equivalent noise-kernel form is
\begin{equation}
 \Gamma_{\rm vac}
 =\frac{\kappa^2}{8}\int\dd^4x\,\dd^4x'\,
 \Delta T^{ij}(x)\GH_{ij,kl}(x,x')\Delta T^{kl}(x'),
 \label{eq:variancevac}
\end{equation}
where the Hadamard function is
\begin{equation}
 \GH_{ij,kl}(x,x')
 =\frac{1}{2}\langle0|\{\hat h^{\TT}_{ij}(x),\hat h^{\TT}_{kl}(x')\}|0\rangle.
\end{equation}

The mean number of gravitons radiated when a classical source $T^{\mu\nu}$ acts on the vacuum is $N=\sum_s\int\frac{\dd^3k}{(2\pi)^3}|\alpha_s|^2$. Comparing with Eq.~\eqref{eq:Gammavac0},
we define
\begin{equation}
 N_\Delta\equiv
 \sum_s\int\frac{\dd^3k}{(2\pi)^3}|\Delta\alpha_s(\kk)|^2,
\end{equation}
so that
\begin{equation}
\Gamma_{\rm vac}=\tfrac{1}{2}N_\Delta.
 \label{eq:Nrelation}
\end{equation}
The quantity $N_\Delta$ is the mean number of gravitons in the auxiliary coherent field sourced by $\Delta T^{\mu\nu}$, rather than the difference between the numbers of gravitons emitted along the two physical branches. Equivalently, $N_\Delta$ is the squared phase-space distance between the two branch-conditioned radiation states, so decoherence becomes appreciable precisely when these states become distinguishable. This relation makes precise the intuition that radiative quanta can carry which-path information~\cite{Baym:2009zu,Danielson:2022tdw}.

\begin{figure}[t]
 \centering
 \includegraphics[width=0.85\columnwidth]{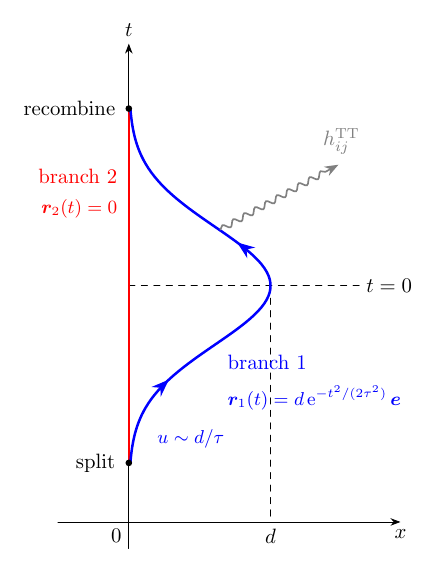}
 \caption{Schematic of the which-path configuration.
 Branch 2 (red) remains at the origin, $\bm r_2(t)=0$, while branch 1 (blue)
 follows the smooth Gaussian excursion
 $\bm r_1(t)=d\,\ee^{-t^2/(2\tau^2)}\bm e$ of
 Eq.~\eqref{eq:gaussiantrajectory}, reaching the maximum separation $d$
 at $t=0$ with characteristic velocity $u\sim d/\tau$. The branches
 coincide asymptotically. The difference source produces TT radiation, and the
 corresponding branch-conditioned field states carry which-path information; tracing out the gravitons
 reduces the interference visibility by $\ee^{-\Gamma}$.}
 \label{fig:whichpath}
\end{figure}

\section{Nonrelativistic quadrupole reduction}
\label{sec:quadrupole}

For slowly moving, compact sources, gravitational radiation is dominated by the time-dependent mass quadrupole moment: mass conservation forbids monopole radiation and momentum conservation forbids dipole radiation, so the quadrupole is the leading radiative multipole~\cite{Weinberg:1972kfs}. This section reduces the master formula~\eqref{eq:masterTT} to that limit.

For a localized source, define the second mass moment in $c=1$ units,
\begin{equation}
 I^{ij}(t)=\int\dd^3x\,x^ix^jT^{00}(t,\xx).
 \label{eq:massmoment}
\end{equation}
Stress-tensor conservation and integration by parts imply
\begin{equation}
 \int\dd^3x\,T^{ij}(t,\xx)=\frac{1}{2}\ddot I^{ij}(t).
 \label{eq:Tijmoment}
\end{equation}
The trace-free mass quadrupole is
\begin{equation}
 Q^{ij}(t)=I^{ij}(t)-\tfrac{1}{3}\delta^{ij}I^{kk}(t).
 \label{eq:STFQ}
\end{equation}
For source size $L$ and frequencies satisfying $\omega L/c\ll1$, the leading term is
\begin{equation}
 \widetilde T^{ij}(\omega,\kk)
 =-\frac{\omega^2}{2}\widetilde I^{ij}(\omega).
 \label{eq:Tlongwave}
\end{equation}
The omitted higher multipoles are suppressed by additional powers of $\omega L/c$. Boundary terms vanish for a closed history or whenever the branch difference and its first derivative decrease sufficiently rapidly at early and late times.

For any symmetric trace-free tensor $A^{ij}$, we have
\begin{equation}
 \int\dd\Omega_{\nn}\,
 \Lambda_{ij,kl}(\nn)A^{ij}A^{kl*}
 =\frac{8\pi}{5}A^{ij}A^{ij*}, 
 \label{eq:angular}
\end{equation}
as proven in Appendix~\ref{app:angular}.
Applying Eqs.~\eqref{eq:Tlongwave} and \eqref{eq:angular} to Eq.~\eqref{eq:masterTT}, noting that the trace part of $\widetilde I^{ij}$ is annihilated by the TT projector, and restoring SI units, we find 
\begin{align}
 \Gamma_{\rm vac}
 &=\frac{G}{10\pi\hbar c^5}
 \int_0^\infty\dd\omega\,\omega^5
 |\Delta\widetilde Q_{ij}(\omega)|^2
 \nonumber\\
 &=\frac{G}{10\pi\hbar c^5}
 \int_0^\infty\frac{\dd\omega}{\omega}
 |\Delta\dddot{\widetilde Q}_{ij}(\omega)|^2,
 \label{eq:quadrupole}
\end{align}
with the shorthand $|\Delta\dddot{\widetilde Q}_{ij}(\omega)|^{2}\equiv\omega^{6}|\Delta\widetilde Q_{ij}(\omega)|^{2}$.
Here $Q_{ij}$ has units of mass times length squared, and $\omega$ is the angular frequency. The associated graviton-number spectrum and diagnostic radiated energy are
\begin{align}
 \frac{\dd N_\Delta}{\dd\omega}
 &=\frac{G}{5\pi\hbar c^5}\omega^5
 |\Delta\widetilde Q_{ij}(\omega)|^2,
 \label{eq:numberspectrum}\\
 E_\Delta
 &=\frac{G}{5\pi c^5}\int_0^\infty\dd\omega\,\omega^6
 |\Delta\widetilde Q_{ij}(\omega)|^2.
 \label{eq:energyspectrum}
\end{align}
$E_\Delta$ is the positive energy radiated by the difference source, not the difference of the energies emitted by the two physical branches. Appendix~\ref{app:energy} verifies the normalization against the Einstein quadrupole formula.

Equation~\eqref{eq:quadrupole} applies strictly to the complete
conserved matter--apparatus source. In the estimates below, however,
we neglect branch-dependent apparatus contributions and use only the
matter quadrupole. For $M\gg m$, the relative recoil correction is of
order $\order(m/M)$; any internal apparatus contribution is
setup-dependent and is not considered here.

For a compact interfering body with branch-dependent displacement $\bm r_b(t)$ relative to the apparatus center of mass, the leading matter contribution is
\begin{equation}
 Q^{\rm m}_{b,ij}(t)
 =m\left[r_{b,i}(t)r_{b,j}(t)
 -\frac{1}{3}\delta_{ij}|\bm r_b(t)|^2\right].
 \label{eq:matterquadrupole}
\end{equation}
The calculations below use Eq.~\eqref{eq:matterquadrupole} directly. Because this quadrupole is even under $\bm r\to-\bm r$, exactly mirror-symmetric branches have $\Delta Q^{\rm m}_{ij}=0$ at this order. Such geometries are controlled by higher-order multipoles.

\section{Gaussian trajectory model}
\label{sec:gaussian}

To obtain a concrete estimate of graviton-induced decoherence, we consider a
simple closed trajectory in which branch $2$ remains at the origin,
whereas branch $1$ undergoes a smooth Gaussian excursion,
\begin{equation}
 \bm r_1(t)
 =d\exp\!\left[-\frac{t^2}{2\tau^2}\right]\bm e,
 \quad
 \bm r_2(t)=0,
 \label{eq:gaussiantrajectory}
\end{equation}
where $\bm e$ is a fixed unit vector, $d$ is the maximum branch
separation, and $\tau$ characterizes the splitting and recombination
time. The two branches coincide in the asymptotic limits
$t\to\pm\infty$; the trajectory profile is shown in
Fig.~\ref{fig:whichpath}.

Using Eq.~\eqref{eq:matterquadrupole} and
$\Delta Q_{ij}=Q_{1,ij}-Q_{2,ij}$, we obtain
\begin{equation}
 \Delta Q_{ij}(t)
 =Q_0\exp\!\left[-\frac{t^2}{\tau^2}\right]S_{ij},
 \quad
 Q_0\equiv md^2,
 \label{eq:gaussianquadrupole}
\end{equation}
where $S_{ij}\equiv e_i e_j-\delta_{ij}/3$ and $S_{ij}S_{ij}=2/3$.
Its Fourier transform is
\begin{equation}
 \Delta\widetilde Q_{ij}(\omega)
 =\sqrt{\pi}\,Q_0\tau
 \exp\!\left[-\frac{(\omega\tau)^2}{4}\right]S_{ij},
 \label{eq:gaussianQFT}
\end{equation}
and hence
\begin{equation}
 |\Delta\widetilde Q_{ij}(\omega)|^2
 =\frac{2\pi}{3}m^2d^4\tau^2
 \exp\!\left[-\frac{(\omega\tau)^2}{2}\right].
 \label{eq:gaussianQFTsq}
\end{equation}

Substituting Eq.~\eqref{eq:gaussianQFTsq} into the quadrupole formula
\eqref{eq:quadrupole} gives
\begin{align}
 \Gamma_{\rm vac}=
 \frac{Gm^2d^4\tau^2}{15\hbar c^5}
 \int_0^\infty \dd\omega\,\omega^5
 \exp\!\left[-\frac{(\omega\tau)^2}{2}\right].
 \label{eq:gaussianGammaOmega}
\end{align}
Introducing $\nu\equiv\omega\tau$, we find
\begin{equation}
 \Gamma_{\rm vac}
 =
 \frac{Gm^2d^4}{15\hbar c^5\tau^4}
 \int_0^\infty \dd\nu\,
 \nu^5\ee^{-\nu^2/2}.
 \label{eq:gaussianGammaNu}
\end{equation}
The remaining integral is
\begin{align}
 \int_0^\infty \dd\nu\,
 \nu^5\ee^{-\nu^2/2}=
 4\int_0^\infty \dd x\,x^2\ee^{-x}=8,
 \label{eq:gaussianintegral}
\end{align}
where $x\equiv\nu^2/2$.
Therefore, the vacuum graviton decoherence exponent is
\begin{equation}
 \Gamma_{\rm vac}
 =
 \frac{8}{15}
 \frac{Gm^2d^4}{\hbar c^5\tau^4}
 =
 \frac{8}{15}
 \left(\frac{m}{\Mpl}\right)^2
 \left(\frac{u}{c}\right)^4,
 \label{eq:gaussgamma}
\end{equation}
where $u\equiv {d}/{\tau}$ and 
\begin{equation}
 \Mpl=\sqrt{\frac{\hbar c}{G}}
 =\SI{2.176434e-8}{\kilogram}
 \label{eq:Planckmass}
\end{equation}
is the Planck mass. 
Equation~\eqref{eq:gaussgamma} makes the two suppression factors explicit:
the interfering mass is typically much smaller than the Planck mass,
$m\ll\Mpl$, and the characteristic velocity scale is nonrelativistic,
$u\ll c$.

The frequency dependence of the decoherence can be characterized by
the contribution per logarithmic frequency interval. From
Eq.~\eqref{eq:gaussianGammaNu}, we obtain
\begin{equation}
 \frac{\dd\Gamma_{\rm vac}}{\dd\nu}
 =
 \frac{Gm^2d^4}{15\hbar c^5\tau^4}
 \nu^5\ee^{-\nu^2/2}.
 \label{eq:GaussNuSpectrum}
\end{equation}
Since
\begin{equation}
 \frac{\dd}{\dd\ln\omega}
 =
 \nu\frac{\dd}{\dd\nu},
\end{equation}
the contribution per logarithmic frequency interval is
\begin{equation}
 \frac{\dd\Gamma_{\rm vac}}{\dd\ln\omega}
 =
 \frac{Gm^2d^4}{15\hbar c^5\tau^4}
 \nu^6\ee^{-\nu^2/2}.
 \label{eq:Gausslogresponse}
\end{equation}
Dividing by the total exponent in Eq.~\eqref{eq:gaussgamma}, we obtain
the normalized logarithmic response
\begin{equation}
 \frac{1}{\Gamma_{\rm vac}}
 \frac{\dd\Gamma_{\rm vac}}{\dd\ln\omega}
 =
 \frac{1}{8}\nu^6\ee^{-\nu^2/2}.
 \label{eq:normalizedGaussSpectrum}
\end{equation}
It is normalized according to
\begin{align}
 \int_0^\infty \dd\ln\omega\,
 \frac{1}{\Gamma_{\rm vac}}
 \frac{\dd\Gamma_{\rm vac}}{\dd\ln\omega}
 &=
 \frac{1}{8}\int_0^\infty\frac{\dd\nu}{\nu}\,
 \nu^6\ee^{-\nu^2/2}
 \nonumber\\
 &=
 \frac{1}{8}\int_0^\infty\dd\nu\,
 \nu^5\ee^{-\nu^2/2}
 =1.
 \label{eq:GaussresponseNormalization}
\end{align}
Thus, Eq.~\eqref{eq:normalizedGaussSpectrum} gives the fraction of the
total decoherence contributed by each logarithmic frequency interval.

\section{Squeezed graviton decoherence}
\label{sec:squeezed}

In this section, we evaluate matter-wave decoherence caused by a squeezed graviton vacuum. Inflation is expected to amplify the vacuum fluctuations of quantum fields and generate primordial gravitational waves in the form of strongly squeezed two-mode states~\cite{Grishchuk:1990bj}, whose enhanced fluctuations can amplify graviton-induced noise and decoherence~\cite{Kanno:2020usf,Kanno:2021gpt}. We first derive the exact decoherence exponent for a general two-mode squeezed vacuum and then specialize to the inflationary case.

\subsection{Squeezed vacuum}
\label{sec:exactsqueezed}

We take the initial graviton state to be a pure two-mode squeezed vacuum that pairs $\kk$ and $-\kk$,
\begin{equation}
 \rho_{\rm g}=|{\rm sq}\rangle\langle{\rm sq}|,
 \quad |{\rm sq}\rangle=\hat S_2[\zeta]|0\rangle,
 \label{eq:sqstate}
\end{equation}
with
\begin{align}
 \hat S_2[\zeta]
 =\exp\!\Bigg\{\frac{1}{2}\sum_s\int\frac{\dd^3k}{(2\pi)^3}
 \Big[&\zeta_s(\kk)\hat a_s^\dagger(\kk)\hat a_s^\dagger(-\kk)
 \nonumber\\[-6pt]
 &-\zeta_s^*(\kk)\hat a_s(\kk)\hat a_s(-\kk)\Big]\Bigg\},
 \label{eq:squeezeoperator}
\end{align}
where we take $\zeta_s(\kk)=r_s(\kk)\ee^{\ii\varphi_s(\kk)}$, $r_s(-\kk)=r_s(\kk)$, and $\varphi_s(-\kk)=\varphi_s(\kk)$. 
The Bogoliubov transformation is
\begin{equation}
 \hat S_2^\dagger\hat a_s(\kk)\hat S_2
 =\cosh r_s\,\hat a_s(\kk)
 +\ee^{\ii\varphi_s}\sinh r_s\,\hat a_s^\dagger(-\kk).
 \label{eq:Bogoliubov}
\end{equation}
The exponent follows from the vacuum formula~\eqref{eq:vacchi} by transferring the squeezing to the displacement,
\begin{equation}
 \langle{\rm sq}|\hat D[\Delta\alpha]|{\rm sq}\rangle
 =\langle0|\hat S_2^\dagger\hat D[\Delta\alpha]\hat S_2|0\rangle.
 \label{eq:transfer}
\end{equation}

Because conjugation by $\hat S_2$ acts linearly on creation and annihilation operators, $\hat S_2^\dagger\hat D[\Delta\alpha]\hat S_2$ is again a displacement operator. Applying Eq.~\eqref{eq:Bogoliubov} and its adjoint to the exponent $\hat X[\Delta\alpha]$ of Eq.~\eqref{eq:Ub} and collecting the coefficient of $\hat a_s^\dagger(\kk)$, we find
\begin{equation}
 \hat S_2^\dagger\hat X[\Delta\alpha]\hat S_2
 =\hat X[\Delta\alpha^{(r)}],
 \label{eq:Xtransformed}
\end{equation}
with the transformed displacement
\begin{align}
 \Delta\alpha_s^{(r)}(\kk)
 &=\cosh r_s(\kk)\Delta\alpha_s(\kk)
 \nonumber\\[-2pt]
 &\quad-\ee^{\ii\varphi_s(\kk)}\sinh r_s(\kk)
 \Delta\alpha_s^*(-\kk).
 \label{eq:alphasqueezed}
\end{align}
Appendix~\ref{app:squeezedchi} performs the same computation explicitly for a single opposite-momentum pair. Applying Eq.~\eqref{eq:vacchi} to the transformed displacement yields the exact exponent
\begin{equation}
 \Gamma_{\rm sq}
 =\frac{1}{2}\sum_s\int\frac{\dd^3k}{(2\pi)^3}
 |\Delta\alpha_s^{(r)}(\kk)|^2,
 \label{eq:Gammasqnorm}
\end{equation}
with
\begin{align}
 |\Delta\alpha_s^{(r)}(\kk)|^2
 &=\cosh^2\!r_s\,|\Delta\alpha_s(\kk)|^2
 +\sinh^2\!r_s\,|\Delta\alpha_s(-\kk)|^2
 \nonumber\\
 &-\sinh(2r_s)
 \operatorname{Re}\!\big[\ee^{-\ii\varphi_s}
 \Delta\alpha_s(\kk)\Delta\alpha_s(-\kk)\big],
 \label{eq:normexpansion}
\end{align}
where $2\cosh r_s\sinh r_s=\sinh(2r_s)$ was used. 
Within the momentum integral, the substitution $\kk\to-\kk$ in the $\sinh^2\!r_s$ term converts $|\Delta\alpha_s(-\kk)|^2$ into $|\Delta\alpha_s(\kk)|^2$, yielding
\begin{align}
 \Gamma_{\rm sq}
 &=\frac{1}{2}\sum_s\int\frac{\dd^3k}{(2\pi)^3}
 \Big\{\cosh(2r_s)|\Delta\alpha_s(\kk)|^2
 \nonumber\\[-2pt]
 &-\sinh(2r_s)\operatorname{Re}\!\left[
 \ee^{-\ii\varphi_s}
 \Delta\alpha_s(\kk)\Delta\alpha_s(-\kk)
 \right]\Big\},
 \label{eq:Gammasq}
\end{align}
where we used $\cosh^2 r_s+\sinh^2 r_s=\cosh(2r_s)$.

For a compact source in the quadrupole regime, we have $|\Delta\alpha_s(-\kk)|=|\Delta\alpha_s(\kk)|$.
We define the corresponding pair phase $\vartheta_s(\kk)$ by
\begin{equation}
 \Delta\alpha_s(\kk)\Delta\alpha_s(-\kk)
 =
 \ee^{\ii\vartheta_s(\kk)}
 \left|\Delta\alpha_s(\kk)\right|^2.
 \label{eq:pairphase}
\end{equation}
The decoherence exponent can therefore be written as
\begin{align}
\Gamma_{\rm sq}
=\frac{1}{2}\sum_s\int\frac{\dd^3k}{(2\pi)^3}\mathcal W_s(\kk)|\Delta\alpha_s(\kk)|^2,
\label{eq:Gammasqweight}
\end{align}
where $\left|\Delta\alpha_s(\kk)\right|^2$ is weighted by
\begin{align}
&\mathcal W_s(\kk)= 1+2n_s(\kk) \nonumber \\
 &-2\sqrt{n_s(\kk)\bigl[n_s(\kk)+1\bigr]}\,
 \cos\!\left[\vartheta_s(\kk)-\varphi_s(\kk)\right],
 \label{eq:squeezedweight}
\end{align}
with
\begin{align}
\begin{split}
n_s(\kk)&=\sinh^2 r_s(\kk), \quad 
\cosh(2r_s)=1+2n_s,\\ 
&\sinh(2r_s)=
 2\sqrt{n_s(n_s+1)}.
\end{split}
\end{align}
For $r_s=0$, we have $n_s=0$ and $\mathcal W_s=1$, recovering the
vacuum result. Since $-1\leq\cos(\vartheta_s-\varphi_s)\leq1$, the weight satisfies
\begin{equation}
 \ee^{-2r_s(\kk)}
 \leq
 \mathcal W_s(\kk)
 \leq
 \ee^{2r_s(\kk)}.
 \label{eq:squeezedbounds}
\end{equation}
Thus squeezing can suppress or enhance the vacuum decoherence response,
depending on the relative phase between the interferometric response
and the squeezing correlation.

\subsection{Inflationary squeezing and decoherence by primordial gravitons}
\label{sec:inflation}

During primordial inflation, the Bunch--Davies vacuum evolves into a two-mode squeezed state \cite{Grishchuk:1990bj}. Let $\hat a_s(\kk)$ annihilate the early-time vacuum, and let $\hat b_s(\kk)$ annihilate the late-time local adiabatic vacuum. Their relation is
\begin{equation}
 \hat b_s(\kk)=\alpha_k^*\hat a_s(\kk)
 -\beta_k^*\hat a_s^\dagger(-\kk),
 \quad |\alpha_k|^2-|\beta_k|^2=1,
 \label{eq:cosmoBogoliubov}
\end{equation}
with occupation $n_k=|\beta_k|^2=\sinh^2r_k$. The de Sitter-to-radiation example in Appendix~\ref{app:inflationarysqueeze} gives $n_k\propto k^{-4}$ for modes outside the horizon at the transition. In the standard cosmological model, this scaling corresponds to an approximately scale-invariant energy density for modes reentering during radiation domination~\cite{Grishchuk:1990bj,Maggiore:1999vm}.

For primordial graviton modes inside the horizon today, the squeezing phase
accumulated during cosmological propagation varies extremely rapidly
with frequency. Since a laboratory protocol probes a finite frequency
band, this phase oscillates many times across the experimental response
window. The phase-sensitive contribution $2\sqrt{n_s[n_s+1]}
 \cos\!\left[\vartheta_s-\varphi_s\right]$ in
Eq.~\eqref{eq:squeezedweight} therefore largely
cancels in the frequency integral, leaving the phase-insensitive factor
$1+2n_s$, dominated by $2n_s$ for $n_s\gg1$~\cite{Allen:1999xw}, rather than the estimate obtained by assuming a frequency-independent squeezing phase and choosing it to maximize the noise~\cite{Kanno:2021gpt}.
After the phase-sensitive contribution has been averaged out,
Eq.~\eqref{eq:Gammasqweight} reduces to
\begin{align}
 \Gamma_{\rm sq}
 &\simeq
 \frac{1}{2}\sum_s
 \int\frac{\dd^3k}{(2\pi)^3}
 \bigl[1+2n_s(\kk)\bigr]
 \left|\Delta\alpha_s(\kk)\right|^2
 \nonumber\\
 &=
 \Gamma_{\rm vac}+\Gamma_{\rm inf},
 \label{eq:GammaSqPhaseAveraged}
\end{align}
where
\begin{equation}
 \Gamma_{\rm inf}
 \equiv
 \sum_s\int\frac{\dd^3k}{(2\pi)^3}
 n_s(\kk)
 \left|\Delta\alpha_s(\kk)\right|^2
 \label{eq:GammaInflationOccupation}
\end{equation}
is the occupation-induced contribution from the primordial graviton
background.

To express this contribution in terms of the vacuum frequency
response, we decompose the momentum-space measure as
\begin{equation}
 \dd^3k
 =
 k^3\dd\ln\omega\,\dd\Omega_{\hat{\kk}},
 \label{eq:momentumLogMeasure}
\end{equation}
where we used $\omega=ck$ and $\dd\ln\omega=\dd k/k$. The vacuum exponent can
therefore be written as
\begin{align}
 \Gamma_{\rm vac}
 &=
 \int_0^\infty\dd\ln\omega\,
 \frac{1}{2}\sum_s
 \frac{k^3}{(2\pi)^3}
 \int\dd\Omega_{\hat{\kk}}\,
 \left|
 \Delta\alpha_s(\kk)
 \right|^2.
 \label{eq:GammaVacLogIntegral}
\end{align}
Accordingly, the contribution per logarithmic frequency interval is
\begin{equation}
 \frac{\dd\Gamma_{\rm vac}}{\dd\ln\omega}
 =
 \frac{1}{2}\sum_s
 \frac{k^3}{(2\pi)^3}
 \int\dd\Omega_{\hat{\kk}}\,
 \left|
 \Delta\alpha_s(\kk)
 \right|^2.
 \label{eq:VacuumLogSpectrumAlpha}
\end{equation}
For an isotropic and unpolarized primordial background, the occupation
number is independent of the propagation direction and helicity, $n_s(\kk)=n(\omega)$, so Eq.~\eqref{eq:GammaInflationOccupation} becomes
\begin{align}
 \Gamma_{\rm inf}
 &=
 \int_0^\infty\dd\ln\omega\,
 n(\omega)
 \sum_s\frac{k^3}{(2\pi)^3}
 \int\dd\Omega_{\hat{\kk}}\,
 \left|
 \Delta\alpha_s(\kk)
 \right|^2
 \nonumber\\
 &=
 2\int_0^\infty\dd\ln\omega\,
 n(\omega)
 \frac{\dd\Gamma_{\rm vac}}{\dd\ln\omega}.
 \label{eq:GammaInflationGeneral}
\end{align}

The energy density carried by primordial gravitons is
\begin{align}
 \rho_{\rm gw}
 &=\sum_s\int\frac{\dd^3k}{(2\pi)^3}\,
 \hbar\omega_k\,n_s(\kk)
 \nonumber\\
 &=\frac{\hbar}{\pi^2c^3}
 \int_0^\infty\dd\omega\,
 \omega^3\,n(\omega),
 \label{eq:modecount}
\end{align}
where the sum over the two helicities gives a factor of $2$. It follows that
\begin{equation}
 \frac{\dd\rho_{\rm gw}}{\dd\ln\omega}
 =\frac{\hbar\omega^4}{\pi^2c^3}\,n(\omega).
 \label{eq:rhooccupation}
\end{equation}
We define the critical energy density and the 
gravitational-wave energy-density parameter as
\begin{equation}
 \varepsilon_{c,0}
 \equiv\frac{3H_0^2c^2}{8\pi G},
 \quad
 \Omega_{\rm gw}(\omega)
 \equiv\frac{1}{\varepsilon_{c,0}}
 \frac{\dd\rho_{\rm gw}}{\dd\ln\omega}.
 \label{eq:Omegadef}
\end{equation}
The graviton occupation number can then be expressed as
\begin{equation}
 n(\omega)
 =\frac{\pi^2c^3\varepsilon_{c,0}}
 {\hbar\omega^4}\,\Omega_{\rm gw}(\omega).
 \label{eq:OmegaOccupation}
\end{equation}
For $H_0=67.4\,{\rm km\,s^{-1}\,Mpc^{-1}}=2.184\times10^{-18}\,{\rm s}^{-1}$~\cite{Planck:2018vyg}, the critical density is $\varepsilon_{c,0}=7.67\times10^{-10}\,{\rm J\,m^{-3}}$.
Consequently, in terms of the frequency $f=\omega/(2\pi)$, we obtain
\begin{equation}
 n(f)=1.24\times10^{32}
 \left(\frac{\Omega_{\rm gw}(f)}{10^{-16}}\right)
 \left(\frac{\SI{1}{\hertz}}{f}\right)^4.
 \label{eq:numericoccupation}
\end{equation}
The large occupation reflects the small energy $\hbar\omega$ of each low-frequency graviton.

\begin{table*}[t]
\caption{Estimated decoherence exponents for matter-wave experiments. The vacuum column is evaluated using Eq.~\eqref{eq:gaussgamma}, whereas the inflationary column uses Eq.~\eqref{eq:GammaInflationGaussian} with $\Omega_{\rm inf}=10^{-16}$. These values do not model the complete physical configurations of the individual experiments and should be regarded only as order-of-magnitude estimates.}
\label{tab:benchmarks}
\begin{ruledtabular}
\begin{tabular}{lccccc}
System & $m$ [kg] & $d$ [m] & $\tau$ [s] & $\Gamma_{\rm vac}$ & $\Gamma_{\rm inf}$ \\
\colrule
Na-cluster nanoparticle interference \cite{Pedalino_2026}
& $2.8561\times10^{-22}$ & $1.33\times10^{-7}$ & $6.1438\times10^{-3}$ & $2.50\times10^{-81}$ & $1.72\times10^{-55}$ \\
$^{87}$Rb half-meter interferometer \cite{Kovachy:2015xcp}
& $1.4447\times10^{-25}$ & $5.4\times10^{-1}$ & $1$ & $2.47\times10^{-70}$ & $1.20\times10^{-35}$ \\
MAQRO representative target \cite{Kaltenbaek:2023xtz}
& $1.6605\times10^{-17}$ & $1.0\times10^{-7}$ & $1.0\times10^{2}$ & $3.84\times10^{-89}$ & $1.86\times10^{-46}$ \\
Superconducting microsphere \cite{Pino:2018erq}
& $3.3211\times10^{-14}$ & $5.00\times10^{-7}$ & $4.83\times10^{-1}$ & $1.77\times10^{-70}$ & $4.65\times10^{-37}$ \\
Aggressive BMV/QGEM scenario \cite{Bose:2017nin}
& $1.0\times10^{-14}$ & $2.50\times10^{-4}$ & $5.0\times10^{-1}$ & $8.71\times10^{-61}$ & $2.63\times10^{-27}$ \\
\end{tabular}
\end{ruledtabular}
\end{table*}

\begin{figure*}[t]
 \centering
 \includegraphics[width=0.99\textwidth]{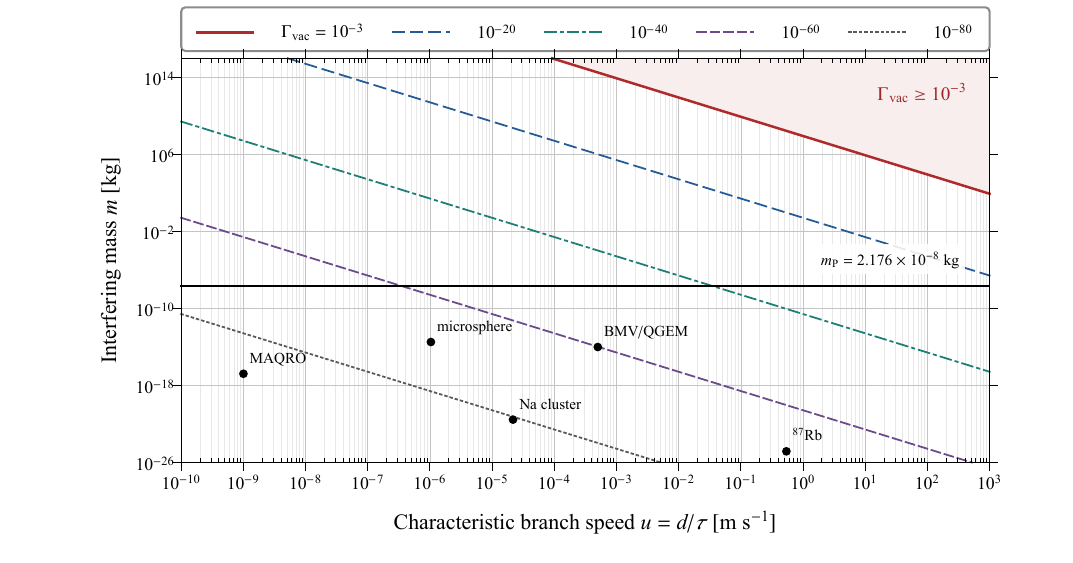}
 \caption{Mass--kinematic-scale dependence of vacuum graviton decoherence in the Gaussian quadrupole model. The descending contours show the interfering mass $m$ required to obtain fixed values of $\Gamma_{\rm vac}$ as a function of the characteristic branch speed $u=d/\tau$, as determined by Eq.~\eqref{eq:gaussgamma}. The thick red contour and the lightly shaded region indicate the optimistic detection threshold $\Gamma_{\rm vac}\geq10^{-3}$. The horizontal black line marks the Planck mass $m_{\rm P}$. The markers represent the experimental proxies listed in Table~\ref{tab:benchmarks}, with their horizontal coordinates evaluated from $u=d/\tau$.}
 \label{fig:parameterplane}
\end{figure*}

For a scale-invariant spectrum $\Omega_{\rm gw}(\omega)=\Omega_{\rm inf}$, 
we insert Eq.~\eqref{eq:OmegaOccupation} at
$\omega=\nu/\tau$,
\begin{equation}
 n(\nu/\tau)
 =\frac{\pi^2c^3\varepsilon_{c,0}\,\Omega_{\rm inf}\,\tau^4}
 {\hbar\,\nu^4}.
 \label{eq:nofnu}
\end{equation}
Substituting this expression, together with the logarithmic
response~\eqref{eq:Gausslogresponse}, into
Eq.~\eqref{eq:GammaInflationGeneral}, we obtain
\begin{align}
 \Gamma_{\rm inf}
 &=\frac{2\pi^2Gm^2d^4\varepsilon_{c,0}\Omega_{\rm inf}}
 {15\,\hbar^2c^2}
 \int_0^\infty\dd\nu\,\nu\,\ee^{-\nu^2/2}
 \nonumber\\
 &=\frac{2\pi^2}{15}
 \frac{Gm^2d^4\varepsilon_{c,0}}{\hbar^2c^2}\Omega_{\rm inf},
 \label{eq:Gammainfsteps}
\end{align}
where the remaining integral equals unity. 
Eliminating $\varepsilon_{c,0}$ with the relation
$G\varepsilon_{c,0}/c^2=3H_0^2/(8\pi)$, which follows from
Eq.~\eqref{eq:Omegadef}, yields
\begin{equation}
 \Gamma_{\rm inf}=\frac{\pi}{20}\Omega_{\rm inf}
 \left(\frac{m d^2H_0}{\hbar}\right)^2.
 \label{eq:GammaInflationGaussian}
\end{equation}

For high-frequency modes reentering well above the electroweak scale during radiation domination, the standard
radiation-era transfer relation for a nearly scale-invariant primordial tensor spectrum gives~\cite{Maggiore:1999vm,Watanabe:2006qe}
\begin{equation}
 \Omega_{\rm inf}
 \simeq\mathcal G\frac{\Omega_{r,0}}{24}r_TA_s,
 \quad
 \mathcal G\simeq0.39.
 \label{eq:inflationaryplateau}
\end{equation}
Here $\mathcal G$ accounts for changes in the effective relativistic degrees of freedom in a standard thermal history.
Under these assumptions, using $\Omega_{r,0}\simeq9.2\times10^{-5}$,
$A_s\simeq2.1\times10^{-9}$, and the BK18 upper limit $r_T\equiv r_{0.05}<0.036$,
we obtain an optimistic plateau of order
$\Omega_{\rm inf}\lesssim1.1\times10^{-16}$
\cite{Planck:2018vyg,BICEP:2021xfz}. In the numerical estimates below, we use $10^{-16}$ as a favorable benchmark.

\section{Experimental estimates}
\label{sec:experiment}

We now evaluate the two decoherence exponents using parameters representative of current and future matter-wave experiments. Specifically, we compare Eq.~\eqref{eq:gaussgamma}, which describes matter decoherence induced by vacuum gravitons, with Eq.~\eqref{eq:GammaInflationGaussian}, which gives the contribution from gravitons of inflationary origin. For the latter, we set $\Omega_{\rm inf}=10^{-16}$ and $H_0=67.4\,{\rm km\,s^{-1}\,Mpc^{-1}}=2.184\times10^{-18}\,{\rm s}^{-1}$. The quantities $m$, $d$, and $\tau$ in Table~\ref{tab:benchmarks} define the effective quadrupole scale $Q_0=md^2$ and the response time. For the nanoparticle, superconducting-microsphere, and BMV entries, the quoted $\tau$ is respectively one grating-to-grating flight time, the stage time $t_3$, and the splitting time $\tau_{\rm acc}$, rather than the total protocol duration. These representative values provide optimistic order-of-magnitude estimates based on matter-wave experiments that have been demonstrated or proposed; they do not reproduce the complete physical setup of each experiment. In particular, $d$ does not necessarily correspond to the actual arm spacing in the cited experiments. The fundamental constants $G$, $c$, and $\hbar$, as well as the atomic-mass conversion, used in the numerical estimates are taken from
the 2022 CODATA constants~\cite{Mohr:2024kco}. As a concrete choice, we adopt
$\Gamma_{\rm det}=10^{-3}$ as an optimistic benchmark
for detecting decoherence, which corresponds to a $0.1\%$ reduction in visibility.
Although the exact detection limit depends on the experiment,
changing this benchmark does not affect the conclusion.

As an explicit BMV splitting-stage proxy, consider $m=10^{-14}\,{\rm kg}$, $d=2.5\times10^{-4}\,{\rm m}$, and $\tau=0.5\,{\rm s}$~\cite{Bose:2017nin}. The dimensionless ratios are then $m/\Mpl=4.60\times10^{-7}$ and $u/c=(d/\tau)/c=1.67\times10^{-12}$, and Eq.~\eqref{eq:gaussgamma} yields
\begin{equation}
 \Gamma_{\rm vac}
 =8.7\times10^{-61}.
 \label{eq:workedvac}
\end{equation}
For the contribution from gravitons of inflationary origin, $md^2=6.25\times10^{-22}\,{\rm kg\,m^2}$ gives $md^2H_0/\hbar=1.29\times10^{-5}$, and Eq.~\eqref{eq:GammaInflationGaussian} yields
\begin{equation}
 \Gamma_{\rm inf}
 =2.6\times10^{-27}.
 \label{eq:workedinf}
\end{equation}
Inflationary squeezing enhances the decoherence induced by primordial gravitons by a factor of approximately $10^{40}$ relative to the vacuum contribution. Nevertheless, even adopting the optimistic benchmark $\Gamma_{\rm det}=10^{-3}$ for detectable decoherence, the predicted effect remains far below experimental sensitivity. We therefore conclude that observing graviton-induced decoherence in present or foreseeable matter-wave interferometry experiments is extremely challenging.

\section{Discussion and conclusions}
\label{sec:discussion}

We have discussed graviton-induced which-path decoherence in matter-wave interferometry. 
Because the matter source couples linearly to the graviton field,
the branch-conditioned evolution can be solved exactly in terms of a coherent
displacement, and the surviving coherence is the characteristic function
of the initial graviton state evaluated at the displacement difference~\eqref{eq:exactD}. In vacuum, this reduces to Eq.~\eqref{eq:Nrelation}.

For a smooth Gaussian closed history the quadrupole reduction gives
$\Gamma_{\rm vac}=(8/15)(m/\Mpl)^2(u/c)^4$,
making explicit the two suppression factors $m\ll\Mpl$ and $u\ll c$; the profile
enters only through an $\mathcal O(1)$ coefficient. For a general squeezed
vacuum the exact response is reweighted between $e^{-2r_s}$ and $e^{2r_s}$, and for the inflationary background the
oscillating squeezing phase averages out, leaving
$\Gamma_{\rm inf}=(\pi/20)\,\Omega_{\rm inf}(md^2H_0/\hbar)^2$. 
Numerically, the vacuum exponent ranges from $10^{-89}$ to $10^{-61}$. Even an optimistic inflationary squeezing with $\Omega_{\rm inf}=10^{-16}$ enhances the effect by at most $\sim10^{40}$, yielding only $\sim3\times10^{-27}$.

We emphasize that the effect computed here is the decoherence due to
\emph{radiative} gravitons, and is physically distinct from the static,
longitudinal (Newtonian) interaction responsible for gravitationally
induced entanglement~\cite{Bose:2017nin,Marletto:2017kzi}; 
the latter is generated by
the constraint sector rather than by on-shell radiation. Our central
conclusion is that graviton-induced which-path decoherence is entirely
negligible for current and foreseeable matter-wave interferometers, so
that radiative gravitons would neither obstruct nor assist efforts to probe the
quantum nature of gravity in the laboratory.

The present analysis is restricted to freely propagating gravitons in
Minkowski spacetime. Near a Killing horizon, by contrast, the mean number
of emitted soft gravitons grows linearly with the time for which a
superposition is held open, and the decoherence is
unbounded~\cite{Danielson:2022tdw,Danielson:2022sga,Danielson:2024yru}. Extending the exact
finite-time framework developed here to such settings, and to more
general nonclassical graviton states, is a natural direction for
future work.

\begin{acknowledgments}
H. M. thanks Toshinori Hayashi, Sugumi Kanno, Shinji Mukohyama, Atsushi Naruko, Kota Numajiri, Hidetoshi Omiya, and Atsushi Taruya for helpful discussions on the detection of gravitational waves using an atomic interferometer and gravitational decoherence. This work is supported by JSPS KAKENHI Grant No. JP23K13100.
\end{acknowledgments}

\section*{Data availability}
No data were created or analyzed in this theoretical study.

\appendix

\section{Angular average}
\label{app:angular}

Let $A_{ij}$ be symmetric and trace-free: $A_{ij}=A_{ji}$ and $A_{ii}=0$. Contracting Eq.~\eqref{eq:TTprojector} with $A_{ij}A^*_{kl}$ gives
\begin{equation}
 \Lambda_{ij,kl}A_{ij}A^*_{kl}
 =A_{ij}A^*_{ij}
 -2\hat k_i\hat k_kA_{ij}A^*_{kj}
 +\tfrac{1}{2}|\hat k_i\hat k_jA_{ij}|^2.
\end{equation}
The angular averages are
\begin{align}
 \int\dd\Omega\,\hat k_i\hat k_j
 &=\frac{4\pi}{3}\delta_{ij},\\
 \int\dd\Omega\,\hat k_i\hat k_j\hat k_k\hat k_l
 &=\frac{4\pi}{15}
 (\delta_{ij}\delta_{kl}+\delta_{ik}\delta_{jl}+\delta_{il}\delta_{jk}).
\end{align}
Using $A_{ii}=0$ gives
\begin{align}
 \int\dd\Omega\,\hat k_i\hat k_kA_{ij}A^*_{kj}
 &=\frac{4\pi}{3}A_{ij}A^*_{ij},\\
 \int\dd\Omega\,|\hat k_i\hat k_jA_{ij}|^2
 &=\frac{8\pi}{15}A_{ij}A^*_{ij}.
\end{align}
Thus,
\begin{equation}
 \int\dd\Omega\,\Lambda_{ij,kl}A_{ij}A^*_{kl}
 =\frac{8\pi}{5}A_{ij}A^*_{ij},
\end{equation}
which is Eq.~\eqref{eq:angular}.

\section{Einstein quadrupole formula}
\label{app:energy}

For an on-shell Fourier component of a conserved source, stress-energy conservation gives $k^\mu\widetilde T_{\mu\nu}(\omega_k,\kk)=0$, with $k^\mu=(\omega_k,\kk)$. The TT-projected contraction can then be written covariantly~\cite{Weinberg:1972kfs}:
\begin{equation}
 \Lambda_{ij,kl}(\nn)\,
 \widetilde T^{ij}\widetilde T^{kl*}
 =\widetilde T^*_{\mu\nu}\widetilde T^{\mu\nu}
 -\tfrac{1}{2}\big|\widetilde T^{\lambda}{}_{\lambda}\big|^2.
 \label{eq:covariantidentity}
\end{equation}
In natural units, the coherent radiation energy is
\begin{align}
 E
 &=\sum_s\int\frac{\dd^3k}{(2\pi)^3}\omega_k|\alpha_s(\kk)|^2
 \nonumber\\
 &=\frac{G}{2\pi^2}\int_0^\infty\dd\omega\,\omega^2
 \int\dd\Omega\,
 \left(\widetilde T^*_{\mu\nu}\widetilde T^{\mu\nu}
 -\tfrac{1}{2}|\widetilde T^\lambda{}_{\lambda}|^2\right),
 \label{eq:energyapp}
\end{align}
Using Eq.~\eqref{eq:covariantidentity}, Eq.~\eqref{eq:energyapp} reproduces Weinberg's classical radiation spectrum in the Fourier convention of Eq.~\eqref{eq:FTdef}~\cite{Weinberg:1972kfs}. Restoring SI units, the quadrupole limit is
\begin{equation}
 E=\frac{G}{5\pi c^5}\int_0^\infty\dd\omega\,\omega^6
 |\widetilde Q_{ij}(\omega)|^2.
\end{equation}
Parseval's identity gives
\begin{equation}
 \int_0^\infty\dd\omega\,\omega^6|\widetilde Q_{ij}(\omega)|^2
 =\pi\int_{-\infty}^{\infty}\dd t\,\dddot Q_{ij}(t)\dddot Q_{ij}(t),
\end{equation}
and hence
\nopagebreak[4]
\begin{equation}
 E=\frac{G}{5c^5}\int\dd t\,\dddot Q_{ij}\dddot Q_{ij},
\end{equation}
which is the Einstein quadrupole formula. Equation~\eqref{eq:Nrelation} also gives $\Gamma_{\rm vac}=\frac{1}{2}\int_0^\infty\dd\omega\,(\hbar\omega)^{-1}\dd E_\Delta/\dd\omega$.

\section{de Sitter-to-radiation transition}
\label{app:inflationarysqueeze}

For either tensor polarization on a spatially flat FLRW background, we define $v_k=a h_k$, absorbing any time-independent normalization into $h_k$. In conformal time and units $c=1$, $v_k$ obeys
\begin{equation}
 v_k''+\left(k^2-\frac{a''}{a}\right)v_k=0.
\end{equation}
For simplicity, we consider an instantaneous transition from exact de Sitter expansion to radiation domination at $\eta=\eta_1$,
\begin{equation}
 a(\eta)=
 \begin{cases}
 -[H(\eta-2\eta_1)]^{-1}, & \eta<\eta_1,\\[2pt]
 \eta/(H\eta_1^2), & \eta>\eta_1.
 \end{cases}
\end{equation}
Here $H>0$ is the de Sitter Hubble rate and $\eta_1>0$.

We take the following mode function for de Sitter and radiation domination, respectively
\begin{align}
 f_k(\eta)
 &=\frac{1}{\sqrt{2k}}
 \left(1-\frac{\ii}{k(\eta-2\eta_1)}\right)
 \ee^{-\ii k(\eta-2\eta_1)},
 \nonumber\\
 g_k(\eta)&=\frac{1}{\sqrt{2k}}\ee^{-\ii k\eta}.
\end{align}
The continuation of $f_k$ into the radiation era can be expanded in the $g_k$ basis as
\begin{equation}
 f_k(\eta)=\alpha_k^*g_k(\eta)-\beta_k g_k^*(\eta),
\end{equation}
Matching $f_k$ and $f_k'$ at $\eta_1$ yields
\begin{align}
 \alpha_k^*&=\left(1-\frac{1}{2k^2\eta_1^2}+\frac{\ii}{k\eta_1}\right)
 \ee^{2\ii k\eta_1},
 \nonumber\\
 \beta_k&=-\frac{1}{2k^2\eta_1^2}.
\end{align}
These coefficients satisfy $|\alpha_k|^2-|\beta_k|^2=1$, and the occupation number is
\begin{equation}
 n_k=|\beta_k|^2=\frac{1}{4k^4\eta_1^4}.
\end{equation}
The equal-time mode amplitude is
\begin{equation}
 |f_k(\eta)|^2
 =\frac{1}{2k}\left[
 1+2n_k-2\operatorname{Re}(\alpha_k^*\beta_k^*\ee^{-2\ii k\eta})
 \right].
\end{equation}
For subhorizon modes, the last term is the anomalous squeezed correlation whose phase evolves as $2k\eta$.

\section{Two-mode squeezed-vacuum displacement}
\label{app:squeezedchi}

For a fixed polarization $s$ and one opposite-momentum pair, set $\hat a\equiv\hat a_s(\kk)$ and $\hat b\equiv\hat a_s(-\kk)$, and we write
\begin{equation}
 \hat S_2(z)=\exp(z\hat a^\dagger\hat b^\dagger-z^*\hat a\hat b),
 \qquad z=r\ee^{\ii\varphi}.
\end{equation}
Here $r\geq0$, and $\varphi$ is the squeezing phase. With this sign convention, the Bogoliubov transformations are
\begin{align}
 \hat S_2^\dagger\hat a\hat S_2
 &=\hat a\cosh r+\ee^{\ii\varphi}\hat b^\dagger\sinh r,
 \nonumber\\
 \hat S_2^\dagger\hat b\hat S_2
 &=\hat b\cosh r+\ee^{\ii\varphi}\hat a^\dagger\sinh r.
\end{align}
Define the two-mode displacement by
\begin{equation}
 \hat D(\beta_a,\beta_b)
 =\exp(\beta_a\hat a^\dagger-\beta_a^*\hat a
 +\beta_b\hat b^\dagger-\beta_b^*\hat b),
\end{equation}
Conjugating by $\hat S_2$ gives
\begin{equation}
 \hat S_2^\dagger\hat D(\beta_a,\beta_b)\hat S_2
 =\hat D(\beta_a',\beta_b'),
\end{equation}
where
\begingroup
\setlength{\abovedisplayskip}{10pt}
\begin{align}
 \beta_a'&=\beta_a\cosh r-\ee^{\ii\varphi}\beta_b^*\sinh r,
 \nonumber\\
 \beta_b'&=\beta_b\cosh r-\ee^{\ii\varphi}\beta_a^*\sinh r.
\end{align}
\endgroup
For $|{\rm sq}\rangle=\hat S_2(z)|0\rangle$, the vacuum characteristic function gives
\begin{align}
 \langle{\rm sq}|\hat D(\beta_a,\beta_b)|{\rm sq}\rangle
 &=\exp(-\Gamma_{\rm pair}),
 \nonumber\\
 \Gamma_{\rm pair}
 &=\frac{1}{2}\cosh(2r)(|\beta_a|^2+|\beta_b|^2)
 \nonumber\\[-2pt]
 &\quad-\sinh(2r)\operatorname{Re}
 [\ee^{-\ii\varphi}\beta_a\beta_b].
\end{align}
After restoring the polarization label, summing $\Gamma_{\rm pair}$ over independent opposite-momentum pairs is equivalent to integrating over all $\kk$ with an overall factor of $1/2$, reproducing Eqs.~\eqref{eq:Gammasqnorm} and \eqref{eq:Gammasq}.

\bibliographystyle{JHEP}
\bibliography{References}

@article{Joos:1984uk,
    author = "Joos, E. and Zeh, H. D.",
    title = "{The Emergence of classical properties through interaction with the environment}",
    doi = "10.1007/BF01725541",
    journal = "Z. Phys. B",
    volume = "59",
    pages = "223--243",
    year = "1985"
}

@article{Zurek:2003zz,
    author = "Zurek, Wojciech Hubert",
    title = "{Decoherence, einselection, and the quantum origins of the classical}",
    eprint = "quant-ph/0105127",
    archivePrefix = "arXiv",
    doi = "10.1103/RevModPhys.75.715",
    journal = "Rev. Mod. Phys.",
    volume = "75",
    pages = "715--775",
    year = "2003"
}

@book{Breuer:2002pc,
    author = "Breuer, H. P. and Petruccione, F.",
    title = "{The theory of open quantum systems}",
    publisher = "Oxford University Press",
    address = "Oxford",
    year = "2002"
}

@article{Anastopoulos:2013zya,
    author = "Anastopoulos, C. and Hu, B. L.",
    title = "{A Master Equation for Gravitational Decoherence: Probing the Textures of Spacetime}",
    eprint = "1305.5231",
    archivePrefix = "arXiv",
    primaryClass = "gr-qc",
    doi = "10.1088/0264-9381/30/16/165007",
    journal = "Class. Quant. Grav.",
    volume = "30",
    pages = "165007",
    year = "2013"
}

@article{Blencowe:2012mp,
    author = "Blencowe, M. P.",
    title = "{Effective Field Theory Approach to Gravitationally Induced Decoherence}",
    eprint = "1211.4751",
    archivePrefix = "arXiv",
    primaryClass = "quant-ph",
    doi = "10.1103/PhysRevLett.111.021302",
    journal = "Phys. Rev. Lett.",
    volume = "111",
    number = "2",
    pages = "021302",
    year = "2013"
}

@article{Oniga:2015lro,
    author = "Oniga, Teodora and Wang, Charles H.-T.",
    title = "{Quantum gravitational decoherence of light and matter}",
    eprint = "1511.06678",
    archivePrefix = "arXiv",
    primaryClass = "quant-ph",
    doi = "10.1103/PhysRevD.93.044027",
    journal = "Phys. Rev. D",
    volume = "93",
    pages = "044027",
    year = "2016"
}

@article{Carney:2018ofe,
    author = "Carney, Daniel and Stamp, Philip C. E. and Taylor, Jacob M.",
    title = "{Tabletop experiments for quantum gravity: a user{\textquoteright}s manual}",
    eprint = "1807.11494",
    archivePrefix = "arXiv",
    primaryClass = "quant-ph",
    doi = "10.1088/1361-6382/aaf9ca",
    journal = "Class. Quant. Grav.",
    volume = "36",
    number = "3",
    pages = "034001",
    year = "2019"
}

@article{Baym:2009zu,
    author = "Baym, Gordon and Ozawa, Tomoki",
    title = "{Two-slit diffraction with highly charged particles: Niels Bohr's consistency argument that the electromagnetic field must be quantized}",
    eprint = "0902.2615",
    archivePrefix = "arXiv",
    primaryClass = "quant-ph",
    doi = "10.1073/pnas.0813239106",
    journal = "Proc. Nat. Acad. Sci.",
    volume = "106",
    pages = "3035--3040",
    year = "2009"
}

@article{Riedel:2013yca,
    author = "Riedel, C. Jess",
    title = "{Evidence for gravitons from decoherence by bremsstrahlung}",
    eprint = "1310.6347",
    archivePrefix = "arXiv",
    primaryClass = "quant-ph",
    month = "10",
    year = "2013"
}

@article{Danielson:2022tdw,
    author = "Danielson, Daine L. and Satishchandran, Gautam and Wald, Robert M.",
    title = "{Black holes decohere quantum superpositions}",
    eprint = "2205.06279",
    archivePrefix = "arXiv",
    primaryClass = "hep-th",
    doi = "10.1142/S0218271822410036",
    journal = "Int. J. Mod. Phys. D",
    volume = "31",
    number = "14",
    pages = "2241003",
    year = "2022"
}

@article{Danielson:2022sga,
    author = "Danielson, Daine L. and Satishchandran, Gautam and Wald, Robert M.",
    title = "{Killing horizons decohere quantum superpositions}",
    eprint = "2301.00026",
    archivePrefix = "arXiv",
    primaryClass = "hep-th",
    doi = "10.1103/PhysRevD.108.025007",
    journal = "Phys. Rev. D",
    volume = "108",
    number = "2",
    pages = "025007",
    year = "2023"
}

@article{Danielson:2024yru,
    author = "Danielson, Daine L. and Satishchandran, Gautam and Wald, Robert M.",
    title = "{Local description of decoherence of quantum superpositions by black holes and other bodies}",
    eprint = "2407.02567",
    archivePrefix = "arXiv",
    primaryClass = "hep-th",
    doi = "10.1103/PhysRevD.111.025014",
    journal = "Phys. Rev. D",
    volume = "111",
    number = "2",
    pages = "025014",
    year = "2025"
}

@article{Feynman:1963fq,
    author = "Feynman, R. P. and Vernon, Jr., F. L.",
    title = "{The Theory of a general quantum system interacting with a linear dissipative system}",
    doi = "10.1016/0003-4916(63)90068-X",
    journal = "Annals Phys.",
    volume = "24",
    pages = "118--173",
    year = "1963"
}

@article{Kanno:2020usf,
    author = "Kanno, Sugumi and Soda, Jiro and Tokuda, Junsei",
    title = "{Noise and decoherence induced by gravitons}",
    eprint = "2007.09838",
    archivePrefix = "arXiv",
    primaryClass = "hep-th",
    reportNumber = "OU-HET-1065, KOBE-COSMO-20-12",
    doi = "10.1103/PhysRevD.103.044017",
    journal = "Phys. Rev. D",
    volume = "103",
    number = "4",
    pages = "044017",
    year = "2021"
}

@article{Kanno:2021gpt,
    author = "Kanno, Sugumi and Soda, Jiro and Tokuda, Junsei",
    title = "{Indirect detection of gravitons through quantum entanglement}",
    eprint = "2103.17053",
    archivePrefix = "arXiv",
    primaryClass = "gr-qc",
    reportNumber = "KOBE-COSMO-21-06",
    doi = "10.1103/PhysRevD.104.083516",
    journal = "Phys. Rev. D",
    volume = "104",
    number = "8",
    pages = "083516",
    year = "2021"
}

@article{Kanno:2025how,
    author = "Kanno, Sugumi and Soda, Jiro and Taniguchi, Akira",
    title = "{Coherent State Description of Gravitational Waves from Binary Black Holes}",
    eprint = "2508.17947",
    archivePrefix = "arXiv",
    primaryClass = "gr-qc",
    reportNumber = "YITP-25-129, KOBE-COSMO-25-16",
    doi = "10.1103/kv1t-j27m",
    journal = "Phys. Rev. Lett.",
    volume = "136",
    number = "6",
    pages = "061404",
    year = "2026"
}

@article{Kanno:2025fpz,
    author = "Kanno, Sugumi and Soda, Jiro and Taniguchi, Akira",
    title = "{Binary gravitational waves as probes of quantum graviton states}",
    eprint = "2510.23326",
    archivePrefix = "arXiv",
    primaryClass = "gr-qc",
    reportNumber = "YITP-25-169, KOBE-COSMO-25-17",
    doi = "10.1103/yp4j-v7lg",
    journal = "Phys. Rev. D",
    volume = "113",
    number = "12",
    pages = "123542",
    year = "2026"
}

@article{Dorlis:2025zzz,
    author = "Dorlis, Panagiotis and Mavromatos, Nick E. and Sarkar, Sarben and Vlachos, Sotirios-Neilos",
    title = "{Superradiant Axionic Black-Hole Clouds as Seeds for Graviton Squeezing}",
    eprint = "2507.01689",
    archivePrefix = "arXiv",
    primaryClass = "gr-qc",
    reportNumber = "KCL-PH-TH/2025-15",
    doi = "10.1103/9crd-zj6l",
    journal = "Phys. Rev. Lett.",
    volume = "135",
    number = "15",
    pages = "151501",
    year = "2025"
}

@article{Dorlis:2025amf,
    author = "Dorlis, Panagiotis and Mavromatos, Nick E. and Sarkar, Sarben and Vlachos, Sotirios-Neilos",
    title = "{Squeezed gravitons from superradiant axion fields around rotating black holes}",
    eprint = "2507.23475",
    archivePrefix = "arXiv",
    primaryClass = "gr-qc",
    reportNumber = "KCL-PH-TH/2025-26",
    doi = "10.1103/t822-86qp",
    journal = "Phys. Rev. D",
    volume = "113",
    number = "2",
    pages = "026023",
    year = "2026"
}

@article{Miki:2020hvg,
    author = "Miki, Daisuke and Matsumura, Akira and Yamamoto, Kazuhiro",
    title = "{Entanglement and decoherence of massive particles due to gravity}",
    eprint = "2010.05159",
    archivePrefix = "arXiv",
    primaryClass = "gr-qc",
    doi = "10.1103/PhysRevD.103.026017",
    journal = "Phys. Rev. D",
    volume = "103",
    number = "2",
    pages = "026017",
    year = "2021"
}

@article{Takeda:2026ujz,
    author = "Takeda, Hiroki and Tomizuka, Shogo and Tanaka, Takahiro",
    title = "{Gravitationally Induced Quantum Decoherence of Macroscopic Objects}",
    eprint = "2606.04099",
    archivePrefix = "arXiv",
    primaryClass = "gr-qc",
    month = "6",
    year = "2026"
}

@article{Parikh:2020nrd,
    author = "Parikh, Maulik and Wilczek, Frank and Zahariade, George",
    title = "{The Noise of Gravitons}",
    eprint = "2005.07211",
    archivePrefix = "arXiv",
    primaryClass = "hep-th",
    doi = "10.1142/S0218271820420018",
    journal = "Int. J. Mod. Phys. D",
    volume = "29",
    number = "14",
    pages = "2042001",
    year = "2020"
}

@article{Parikh:2020kfh,
    author = "Parikh, Maulik and Wilczek, Frank and Zahariade, George",
    title = "{Quantum Mechanics of Gravitational Waves}",
    eprint = "2010.08205",
    archivePrefix = "arXiv",
    primaryClass = "hep-th",
    doi = "10.1103/PhysRevLett.127.081602",
    journal = "Phys. Rev. Lett.",
    volume = "127",
    number = "8",
    pages = "081602",
    year = "2021"
}

@article{Parikh:2020fhy,
    author = "Parikh, Maulik and Wilczek, Frank and Zahariade, George",
    title = "{Signatures of the quantization of gravity at gravitational wave detectors}",
    eprint = "2010.08208",
    archivePrefix = "arXiv",
    primaryClass = "hep-th",
    doi = "10.1103/PhysRevD.104.046021",
    journal = "Phys. Rev. D",
    volume = "104",
    number = "4",
    pages = "046021",
    year = "2021"
}

@article{Toros:2020krn,
    author = "Toro{\v{s}}, Marko and Mazumdar, Anupam and Bose, Sougato",
    title = "{Loss of coherence and coherence protection from a graviton bath}",
    eprint = "2008.08609",
    archivePrefix = "arXiv",
    primaryClass = "gr-qc",
    doi = "10.1103/PhysRevD.109.084050",
    journal = "Phys. Rev. D",
    volume = "109",
    number = "8",
    pages = "084050",
    year = "2024"
}

@article{Ford:1982wu,
    author = "Ford, L. H.",
    title = "{Gravitational radiation by quantum systems}",
    reportNumber = "TUTP-82-6",
    doi = "10.1016/0003-4916(82)90115-4",
    journal = "Annals Phys.",
    volume = "144",
    number = "2",
    pages = "238--248",
    year = "1982"
}

@article{Kanno:2018cuk,
    author = "Kanno, Sugumi and Soda, Jiro",
    title = "{Detecting nonclassical primordial gravitational waves with Hanbury-Brown{\textendash}Twiss interferometry}",
    eprint = "1810.07604",
    archivePrefix = "arXiv",
    primaryClass = "hep-th",
    reportNumber = "OU-HET-980, KOBE-COSMO-18-09",
    doi = "10.1103/PhysRevD.99.084010",
    journal = "Phys. Rev. D",
    volume = "99",
    number = "8",
    pages = "084010",
    year = "2019"
}

@article{Kanno:2024gjt,
    author = "Kanno, Sugumi and Matsui, Hiroki and Mukohyama, Shinji",
    title = "{Hanbury-Brown-Twiss interferometry and quantum nature of primordial gravitational waves in Ho{\v{r}}ava-Lifshitz gravity}",
    eprint = "2412.19514",
    archivePrefix = "arXiv",
    primaryClass = "gr-qc",
    reportNumber = "YITP-24-160, IPMU24-0044",
    doi = "10.1103/PhysRevD.111.104077",
    journal = "Phys. Rev. D",
    volume = "111",
    number = "10",
    pages = "104077",
    year = "2025"
}

@article{Grishchuk:1990bj,
    author = "Grishchuk, L. P. and Sidorov, Yu. V.",
    title = "{Squeezed quantum states of relic gravitons and primordial density fluctuations}",
    doi = "10.1103/PhysRevD.42.3413",
    journal = "Phys. Rev. D",
    volume = "42",
    pages = "3413--3421",
    year = "1990"
}

@article{Bose:2017nin,
    author = "Bose, Sougato and Mazumdar, Anupam and Morley, Gavin W. and Ulbricht, Hendrik and Toro{\v{s}}, Marko and Paternostro, Mauro and Geraci, Andrew and Barker, Peter and Kim, M. S. and Milburn, Gerard",
    title = "{Spin Entanglement Witness for Quantum Gravity}",
    eprint = "1707.06050",
    archivePrefix = "arXiv",
    primaryClass = "quant-ph",
    doi = "10.1103/PhysRevLett.119.240401",
    journal = "Phys. Rev. Lett.",
    volume = "119",
    number = "24",
    pages = "240401",
    year = "2017"
}

@article{Marletto:2017kzi,
    author = "Marletto, Chiara and Vedral, Vlatko",
    title = "{Gravitationally Induced Entanglement between Two Massive Particles Is Sufficient Evidence of Quantum Effects in Gravity}",
    eprint = "1707.06036",
    archivePrefix = "arXiv",
    primaryClass = "quant-ph",
    doi = "10.1103/PhysRevLett.119.240402",
    journal = "Phys. Rev. Lett.",
    volume = "119",
    number = "24",
    pages = "240402",
    year = "2017"
}

@article{Matsumura:2020law,
    author = "Matsumura, Akira and Yamamoto, Kazuhiro",
    title = "{Gravity-induced entanglement in optomechanical systems}",
    eprint = "2010.05161",
    archivePrefix = "arXiv",
    primaryClass = "gr-qc",
    doi = "10.1103/PhysRevD.102.106021",
    journal = "Phys. Rev. D",
    volume = "102",
    number = "10",
    pages = "106021",
    year = "2020"
}

@article{Miki:2024qcz,
    author = "Miki, Daisuke and Matsumura, Akira and Yamamoto, Kazuhiro",
    title = "{Feasible generation of gravity-induced entanglement by using optomechanical systems}",
    eprint = "2406.04361",
    archivePrefix = "arXiv",
    primaryClass = "quant-ph",
    doi = "10.1103/PhysRevD.110.024057",
    journal = "Phys. Rev. D",
    volume = "110",
    number = "2",
    pages = "024057",
    year = "2024"
}

@article{Fujita:2023pia,
    author = "Fujita, Tomohiro and Kaku, Youka and Matsumura, Akira and Michimura, Yuta",
    title = "{Inverted oscillators for testing gravity-induced quantum entanglement}",
    eprint = "2308.14552",
    archivePrefix = "arXiv",
    primaryClass = "quant-ph",
    doi = "10.1088/1361-6382/adf0bb",
    journal = "Class. Quant. Grav.",
    volume = "42",
    number = "16",
    pages = "165003",
    year = "2025"
}

@article{Magnus:1954zz,
    author = "Magnus, Wilhelm",
    title = "{On the exponential solution of differential equations for a linear operator}",
    doi = "10.1002/cpa.3160070404",
    journal = "Commun. Pure Appl. Math.",
    volume = "7",
    pages = "649--673",
    year = "1954"
}

@article{Blanes:2008xlr,
    author = "Blanes, S. and Casas, F. and Oteo, J. A. and Ros, J.",
    title = "{The Magnus expansion and some of its applications}",
    doi = "10.1016/j.physrep.2008.11.001",
    journal = "Phys. Rept.",
    volume = "470",
    pages = "151--238",
    year = "2009"
}

@article{Allen:1999xw,
    author = "Allen, Bruce and Flanagan, Eanna E. and Papa, Maria Alessandra",
    title = "{Is the squeezing of relic gravitational waves produced by inflation detectable?}",
    eprint = "gr-qc/9906054",
    archivePrefix = "arXiv",
    reportNumber = "WISC-MILW-99-TH-07",
    doi = "10.1103/PhysRevD.61.024024",
    journal = "Phys. Rev. D",
    volume = "61",
    pages = "024024",
    year = "2000"
}

@article{Maggiore:1999vm,
    author = "Maggiore, Michele",
    title = "{Gravitational wave experiments and early universe cosmology}",
    eprint = "gr-qc/9909001",
    archivePrefix = "arXiv",
    reportNumber = "IFUP-TH-20-99",
    doi = "10.1016/S0370-1573(99)00102-7",
    journal = "Phys. Rept.",
    volume = "331",
    pages = "283--367",
    year = "2000"
}

@article{Watanabe:2006qe,
    author = "Watanabe, Yuki and Komatsu, Eiichiro",
    title = "{Improved Calculation of the Primordial Gravitational Wave Spectrum in the Standard Model}",
    eprint = "astro-ph/0604176",
    archivePrefix = "arXiv",
    doi = "10.1103/PhysRevD.73.123515",
    journal = "Phys. Rev. D",
    volume = "73",
    number = "12",
    pages = "123515",
    year = "2006"
}

@article{Planck:2018vyg,
    author = "Aghanim, N. and others",
    collaboration = "Planck",
    title = "{Planck 2018 results. VI. Cosmological parameters}",
    eprint = "1807.06209",
    archivePrefix = "arXiv",
    primaryClass = "astro-ph.CO",
    doi = "10.1051/0004-6361/201833910",
    journal = "Astron. Astrophys.",
    volume = "641",
    pages = "A6",
    year = "2020",
    note = "[Erratum: Astron.Astrophys. 652, C4 (2021)]"
}

@article{BICEP:2021xfz,
    author = "Ade, P. A. R. and others",
    collaboration = "BICEP, Keck",
    title = "{Improved Constraints on Primordial Gravitational Waves using Planck, WMAP, and BICEP/Keck Observations through the 2018 Observing Season}",
    eprint = "2110.00483",
    archivePrefix = "arXiv",
    primaryClass = "astro-ph.CO",
    doi = "10.1103/PhysRevLett.127.151301",
    journal = "Phys. Rev. Lett.",
    volume = "127",
    number = "15",
    pages = "151301",
    year = "2021"
}

@article{Pedalino_2026,
    author = "Pedalino, Sebastian and Ram{\'i}rez-Galindo, Bruno E. and Ferstl, Richard and Hornberger, Klaus and Arndt, Markus and Gerlich, Stefan",
    title = "{Probing quantum mechanics with nanoparticle matter-wave interferometry}",
    doi = "10.1038/s41586-025-09917-9",
    journal = "Nature",
    volume = "649",
    number = "8098",
    pages = "866--870",
    year = "2026"
}

@article{Kovachy:2015xcp,
    author = "Kovachy, T. and Asenbaum, P. and Overstreet, C. and Donnelly, C. A. and Dickerson, S. M. and Sugarbaker, A. and Hogan, J. M. and Kasevich, M. A.",
    title = "{Quantum superposition at the half-metre scale}",
    doi = "10.1038/nature16155",
    journal = "Nature",
    volume = "528",
    number = "7583",
    pages = "530--533",
    year = "2015"
}

@article{Kaltenbaek:2023xtz,
    author = "Kaltenbaek, Rainer and others",
    title = "{Research campaign: Macroscopic quantum resonators (MAQRO)}",
    doi = "10.1088/2058-9565/aca3cd",
    journal = "Quantum Sci. Technol.",
    volume = "8",
    number = "1",
    pages = "014006",
    year = "2023"
}

@article{Pino:2018erq,
    author = "Pino, H. and Prat-Camps, J. and Sinha, K. and Venkatesh, B. Prasanna and Romero-Isart, O.",
    title = "{On-chip quantum interference of a superconducting microsphere}",
    eprint = "1603.01553",
    archivePrefix = "arXiv",
    primaryClass = "quant-ph",
    doi = "10.1088/2058-9565/aa9d15",
    journal = "Quantum Sci. Technol.",
    volume = "3",
    number = "2",
    pages = "025001",
    year = "2018"
}

@article{Mohr:2024kco,
    author = "Mohr, Peter J. and Newell, David B. and Taylor, Barry N. and Tiesinga, Eite",
    title = "{CODATA recommended values of the fundamental physical constants: 2022}",
    eprint = "2409.03787",
    archivePrefix = "arXiv",
    primaryClass = "hep-ph",
    doi = "10.1103/RevModPhys.97.025002",
    journal = "Rev. Mod. Phys.",
    volume = "97",
    number = "2",
    pages = "025002",
    year = "2025"
}

@book{Weinberg:1972kfs,
    author = "Weinberg, Steven",
    title = "{Gravitation and Cosmology}: {Principles and Applications of the General Theory of Relativity}",
    isbn = "978-0-471-92567-5",
    publisher = "John Wiley and Sons",
    address = "New York",
    year = "1972"
}

\end{document}